\documentclass[pdflatex,sn-nature]{sn-jnl}

\usepackage{graphicx}
\usepackage{multirow}
\usepackage{amsmath,amssymb,amsfonts}
\usepackage{amsthm}
\usepackage{mathrsfs}
\usepackage[title]{appendix}
\usepackage{xcolor}
\usepackage{textcomp}
\usepackage{manyfoot}
\usepackage{booktabs}
\unnumbered

\newcommand{\kms}{km\,s$^{-1}$}
\newcommand{\kmskpc}{km\,s$^{-1}$\,kpc$^{-1}$}
\newcommand{\degGyr}{$^\circ$\,Gyr$^{-1}$}

\begin{document}

\title[Tilting rate of the Milky Way disc]{A direct measurement of the tilting rate of the Milky Way disc}

\author*[1,2]{\fnm{Guillaume F.} \sur{Thomas}}\email{gthomas@iac.es}

\author*[3]{\fnm{Benoit} \sur{Famaey}}\email{benoit.famaey@astro.unistra.fr}

\author[2,1]{\fnm{Alicia} \sur{Rivero}}

\author[2,1]{\fnm{Giuseppina} \sur{Battaglia}}

\author[3]{\fnm{Rodrigo} \sur{Ibata}}

\affil[1]{\orgname{Universidad de La Laguna}, \orgdiv{Dpto. Astrof\'isica}, \orgaddress{\postcode{E-38206}, \city{La Laguna}, \state{Tenerife}, \country{Spain}}}

\affil[2]{\orgname{Instituto de Astrof\'isica de Canarias}, \orgaddress{\postcode{E-38205}, \city{La Laguna}, \state{Tenerife}, \country{Spain}}}

\affil[3]{\orgname{Universit\'e de Strasbourg, CNRS}, \orgdiv{Observatoire Astronomique de Strasbourg, UMR 7550}, \orgaddress{\postcode{F-67000}, \city{Strasbourg}, \country{France}}}

\abstract{The Milky Way disc is typically described, at lowest order, as a flattened rotating stellar component with a fixed spin axis. However, interactions and mergers with smaller galaxies \cite{Eggen1962,Searle1978,White1978,White1991,Springel2005,Bullock2005} are expected to have an impact on the disc properties. Such interactions can leave long-lasting signatures within galactic discs, including warps and corrugation waves that have already been detected in the Milky Way \cite{Gomez2017,Antoja2018,Laporte2019,Poggio2020,Poggio2025}, but also a slow reorientation of the disc known as tilting \cite{Binney1986,Ostriker1989,Huang1997,Sellwood1998,Dodge2023}, which has never been directly measured up to now. Here, using the vertical proper motions of 110 million stars from the Gaia DR3 catalogue \cite{Gaia2021}, we present the first-ever direct measurement of the tilting rate of our Galaxy's disc. The detected signal of 14 $\pm$ 2\degGyr around the Sun--Galactic Centre axis is statistically significant and is consistent with typical values found in cosmological simulations \cite{Gomez2017,Dodge2023,Earp2017,Dillamore2022}. By comparing our measurements with high-resolution tailored simulations, we infer that the observed tilting rate is likely the long-term consequence of a major accretion event that occurred around 8--10 billion years ago. The signal is consistent with the merger of a massive satellite galaxy on a retrograde orbit with a ratio of about 1:4, probably associated with the Gaia--Enceladus/Sausage event \cite{Belokurov2018,Helmi2018,Gallart2019,DiMatteo2019}. These results provide independent evidence for a key episode in the Milky Way's formation history and introduce a new way to reconstruct the assembly histories of disc galaxies from their present-day motions.}

%%\keywords{}

\maketitle

\section{Main}

\begin{figure}[htbp]
\centering
\includegraphics[width=0.9\textwidth]{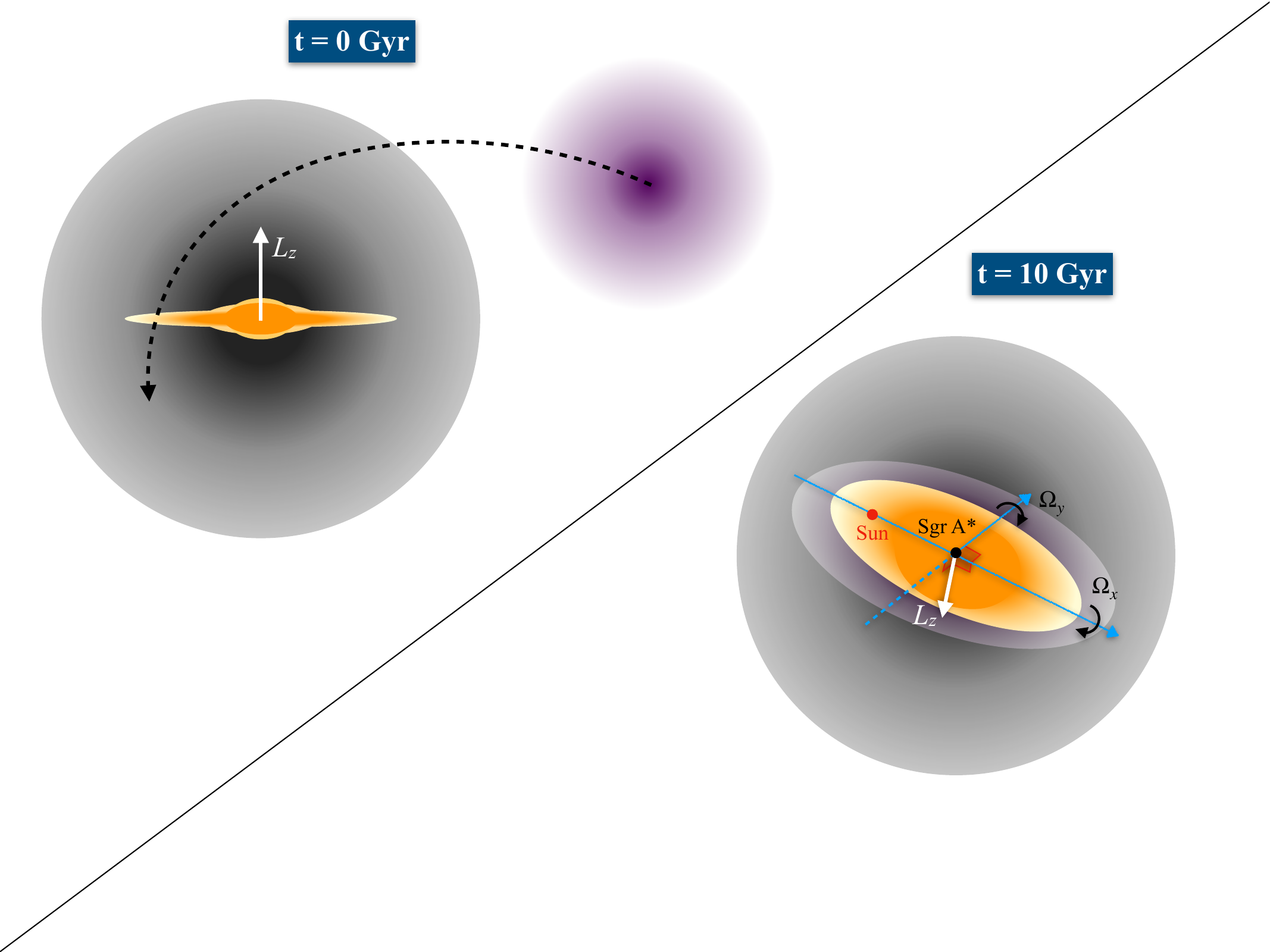}
\caption{\textbf{Schematic representation of the tilting of the Galactic disc induced by a massive merger.} Top, the Galactic disc before its merger with a massive satellite galaxy (purple), such as Gaia--Enceladus/Sausage, whose infall is indicated by the dashed orbit. The white arrow indicates the direction of the disc's vertical angular momentum ($L_z$) at this early time. Bottom, the same system 10~Gyr after the merger, which has reoriented the disc. The angular momentum vector is still slowly tilting towards alignment with the orbital angular momentum direction of the accreted satellite, so that the Galaxy tumbles around this axis. This residual tumbling can be decomposed into two angular velocity components, illustrated for the present-day disc: $\Omega_x$, describing the angular rotation rate about the Sun--Galactic-centre axis (whose Galactic-centre end corresponds to the location of Sgr~A$^*$), and $\Omega_y$, the angular rotation rate about the perpendicular axis, aligned with the disc's direction of rotation.}\label{fig:1}
\end{figure}

A schematic example of the impact of a massive merger on a galactic disc is illustrated in Fig.~\ref{fig:1}, where, in this case, the vertical angular momentum of the disc ($L_z$) tilts by almost 180$^\circ$ over 10 billion years \cite{Dodge2023}. The current tilting rate can be decomposed into two angular velocity components, $\Omega_x$ and $\Omega_y$ in a right-handed Galactocentric reference frame. The former is a rotation about the Sun--Galactic centre axis (hereafter referred to as the X-axis), with the Galactic centre placed at the location of the supermassive black hole Sgr~A$^*$, and the latter a rotation about the axis aligned with the direction of rotation of the Galactic disc (referred to as the Y-axis). In practice, however, measuring both of these rates in the Milky Way directly from six-dimensional phase-space data from the Gaia RVS is complicated due to the fact that the disc is vertically perturbed by corrugation waves that can locally erase any global tilting signal. Hence, any local detection of the tilting rate is not feasible, and one must instead rely either on indirect inferences based on the stellar halo \cite{Nibauer2024,Wang2026} or on the astrometry of the whole Gaia catalogue, which we attempt here for the first time.

Indeed, in the Milky Way, both angular velocities $\Omega_x$ and $\Omega_y$ are in principle recoverable from the measured stellar proper motions \cite{Perryman2014}, since a disc tilt imprints a characteristic signature on the vertical proper motions in Galactic latitude, $\mu_b$, as a function of position on the sky.

The complex vertical corrugation pattern mentioned above is superimposed onto a linear increase of vertical velocities with angular momentum expected from the warp \cite{Schonrich2018,Huang2018}, which makes it very challenging to measure a global $\Omega_y$ tilting rate without strong degeneracies, since the kinematic signature of the warp is comparable or larger to any realistic tilting rate over large regions of the disc. The warp is generally expected to become significant beyond Galactocentric radii of $R > 10$~kpc \cite{Hunt2025}, but the radius at which it first develops remains debated \cite{Drimmel2001,Reyle2009,Amores2017,Cheng2020,Dehnen2023}, and several studies have reported warp-induced kinematic signatures at smaller guiding radii ($R_g \sim 7$~kpc) \cite{Schonrich2018,Huang2018}. Hence, we do not attempt a precise measurement of $\Omega_y$, and only derive instead an upper limit of $|\Omega_y| = 0.1$~\kmskpc\ (see Methods and Fig.~\ref{fig:S1}). However, the associated uncertainties remain too large for this constraint to be considered robust.

However, the situation is much more favourable for $\Omega_x$, which is less affected by the warp and for which the amplitude of the proper-motion signature is distance-independent (see Methods). This limits the impact of the distance uncertainties in measuring it, which are by far the dominant factor for this type of analysis.

\begin{figure}[htbp]
\centering
\includegraphics[width=0.9\textwidth]{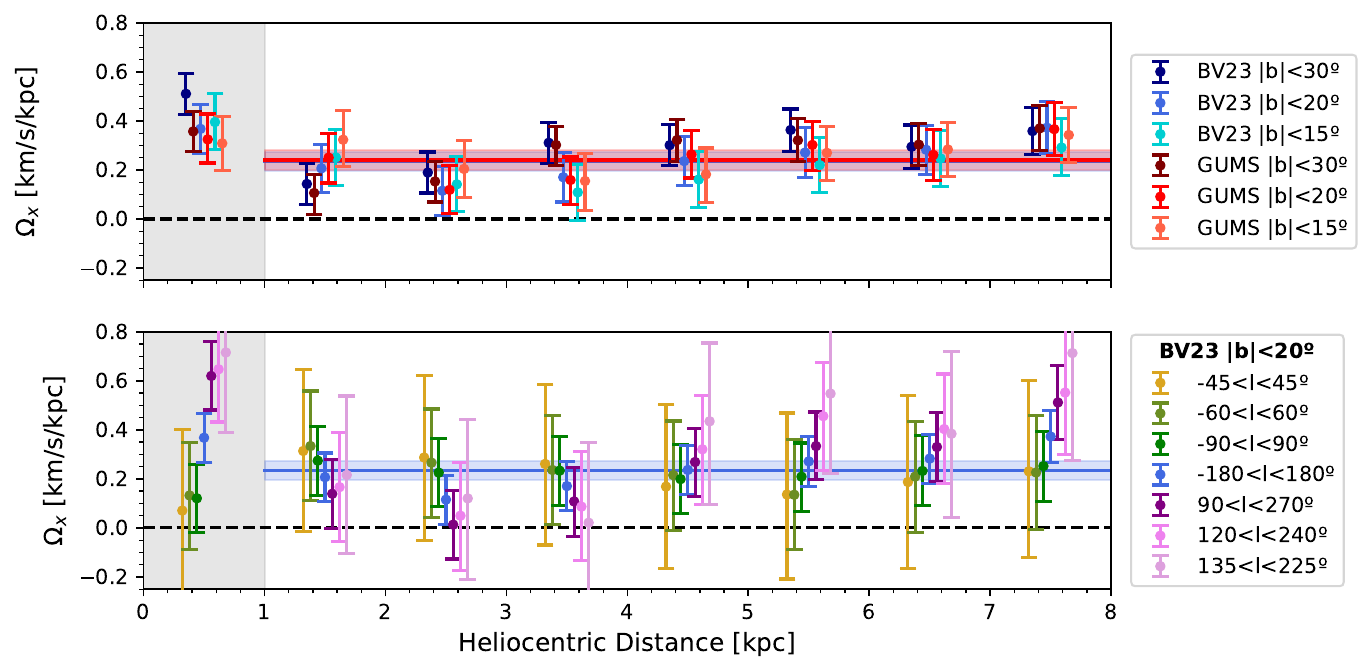}
\caption{\textbf{Present-day tilting rate of the Milky Way disc around the X-axis.} The tilting rate about the Sun--Galactic-centre axis ($\Omega_x$) is shown as a function of heliocentric distance (circles with error bars). The solid lines show $\Omega_x$ measured using all stars between 1 and 8~kpc, with the shaded regions indicating the $1\sigma$ uncertainties. The top panel shows the measurement for different ranges of Galactic latitude. The bottom panel shows $\Omega_x$ measured in the BV23 case within three increasingly wide cones with half-angles of $45^\circ$, $60^\circ$ and $90^\circ$, directed towards the Galactic centre (yellow/green) and Galactic anticentre (pink/purple). In both panels, the dashed line indicates the value expected in the absence of tilting. The error bars indicate $1\sigma$ uncertainties, and the grey shaded region between 0 and 1~kpc corresponds to the distance range over which the measurements are considered unreliable, due to systematics in the Gaia parallaxes.}\label{fig:2}
\end{figure}

We thus determine $\Omega_x$ by comparing the predictions of two independent self-consistent Milky Way models, BV23 \cite{Binney2023} and GUMS \cite{Robin2012} with added tilting (see Methods), to the observed mean proper motion in Galactic latitude, $\mu_b$, on the sky. This quantitative comparison uses more than 110 million stars with astrometric measurements from Gaia, and is performed across the sky and over heliocentric distances up to 8~kpc. The analysis is restricted to Galactic latitudes of $|b| < 30^\circ$ in order for disc stars to dominate the signal, but we also tried lower latitude cuts. As shown in Fig.~\ref{fig:2}, regardless of the adopted Galactic model and distance range considered, we detected a statistically significant tilting rate around the X-axis. Depending on the underlying Milky Way model and the latitude cuts applied, the inferred value of $\Omega_x$ lies between 0.20 and 0.28~\kmskpc\ (12--16\degGyr) when we consider the full distance range, with typical statistical uncertainties of 0.03~\kmskpc. Averaging over all model configurations gives $\Omega_x = 0.24 \pm 0.04$~\kmskpc\ ($14 \pm 2$\degGyr) at one sigma (scatter). When we analyse the tilting rate in 1~kpc distance bin intervals independently, the recovered values range from 0.10 to 0.36~\kmskpc, with typical uncertainties of 0.10~\kmskpc\ (Fig.~\ref{fig:2}, top panel), fully consistent with a distance-independent tilting rate. Moreover, both Milky Way models give highly consistent results.

An analysis of the evidence indeed strongly favours a tilting disc over a static one. For instance, for our fiducial configuration (BV23, $|b| < 20^\circ$, all Galactic longitudes), the Bayes factor for a tilting rate of $\Omega_x = 0.24 \pm 0.04$~\kmskpc\ is $\ln(\mathrm{BF}_{10}) = 12.24$, corresponding to decisive evidence. A similar conclusion is reached with other information criteria ($\Delta\mathrm{BIC} = 24.90$ and $\Delta\mathrm{AIC} = 33.16$).

To test the robustness of this measurement, we repeated the analysis using only stars within progressively narrower cones centred on the Galactic centre and the Galactic anticentre. As expected, the statistical uncertainties grow for smaller opening angles, both because fewer regions of the sky contribute to the fit and because the sensitivity to $\Omega_x$ falls as $\sin(l)$, approaching zero near the Galactic centre and anticentre (see Methods). In all cases, as shown in the lower panel of Fig.~\ref{fig:2}, the inferred value of $\Omega_x$ remains consistent with the global measurement. The only deviation occurs at heliocentric distances of 2--4~kpc towards the anticentre, corresponding to Galactocentric radii of 10--12~kpc. In this region, we attribute the systematically lower value of $\Omega_x$ to the influence of the Galactic warp. Indeed, because the line of nodes of the warp is offset from the Galactic anticentre ($l = 180^\circ$) and precesses with radius \cite{Dehnen2023,Poggio2018,RomeroGomez2019,CabreraGadea2024,Jonsson2024,Poggio2025}, the mean vertical velocity changes from one side of the node to the other, which locally biases the measurement of $\Omega_x$.

Disc tilting is a common feature of Milky Way-like galaxies in cosmological simulations and is most often driven by the accretion of massive satellites, although other mechanisms may also contribute \cite{Gomez2017,Dillamore2022,Wang2026,Bett2012,Dekel2020,Zhu2026}. The tilting can go on for several billion years after the merger, as the debris of the accreted satellite continues to exert a gravitational torque on the Galactic disc \cite{Dodge2023}, producing a slow precession around the angular momentum direction of the merger remnant. The present-day tilting rate we measured in the Milky Way of $14 \pm 2$\degGyr, falls well within the range predicted by these cosmological simulations. It is also well in line with the grossly inferred value from possible indirect signatures in the Palomar 5 stream \cite{Nibauer2024} and somewhat larger than the rough estimates inferred from the amplitude of the Galactic warp \cite{Wang2026}.

\begin{figure}[htbp]
\centering
\includegraphics[width=\textwidth]{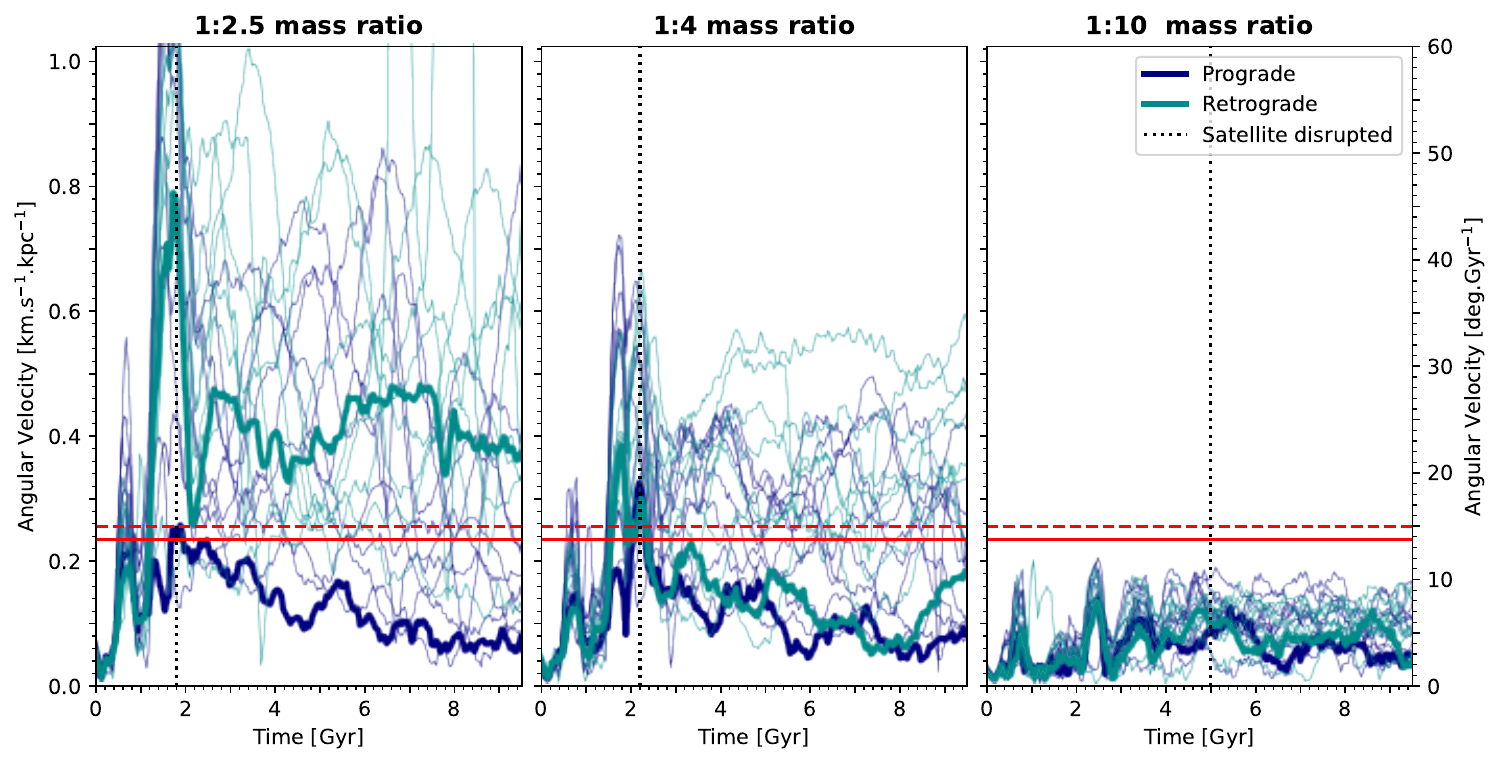}
\caption{\textbf{Evolution of the disc tilting rate induced by a satellite merger.} Each panel shows the evolution of the tilting rate of a Milky Way-like disc following the accretion of a satellite with the mass ratio indicated above. Simulations assume an orbital circularity of 0.5 and an initial position angle of $15^\circ$, and explore different orbital inclinations for both prograde and retrograde orbits with respect to the Galactic disc. The thicker lines highlight the cases of perfectly prograde and retrograde orbits. Tilting rates were computed from the time derivative of the disc orientation after applying a Savitzky--Golay filter to suppress small fluctuations. The solid red line indicates the measured value of $\Omega_x$, while the dashed red line shows the corresponding total tilting rate obtained by adding the tentative value of $\Omega_y$ in quadrature. Vertical dashed lines indicate the time at which the satellite is fully disrupted.}\label{fig:3}
\end{figure}

To better place our measurement in such a context, we performed a suite of tailored collisionless $N$-body simulations of a Milky Way-like galaxy undergoing a merger 10~Gyr ago \cite{Belokurov2018,Helmi2018,Gallart2019,DiMatteo2019}, consistent with the inferred epoch of the accretion of the Gaia--Enceladus/Sausage (G-E/S) galaxy (see Methods), the most important merger in our Galaxy's history. The current Large Magellanic Cloud infall produces, on the other hand, a negligible tilt of the disc incompatible with our measurement \cite{Vasiliev2024}. Assuming that the present-day tilting rate of the Milky Way is only due to the accretion of G-E/S, our measurement favours a retrograde merger with a total mass ratio of around 1:4, in agreement with recent inferences from the rotation signature of old alpha-rich populations \cite{Orkney2026}. Indeed, as can be seen in Fig.~\ref{fig:3}, averaging over the last billion years, a 1:10 merger produces a tilting rate of $\approx 3$\degGyr\ for both prograde and retrograde orbits, with little dependence on the orbital inclination. In contrast, a 1:2.5 merger yields an average tilting rate of 22\degGyr\ for a retrograde orbit, and of 4\degGyr\ for a prograde orbit. For a 1:4 mass ratio, the tilting rate is 4\degGyr\ for a perfect prograde orbit. It is, on average, much closer to our measurement for a retrograde orbit, 9\degGyr. Note that for the 1:2.5 and 1:4 merger ratio, the tilting rate fluctuates strongly depending on the exact orbit inclination, although the tilting rate tends to be higher for retrograde orbits.

This first-ever direct measurement of the present-day tilting rate of the Milky Way disc, using Gaia proper motions, opens a new era in our understanding and modelling of the dynamical state of the Galaxy. It sheds new light on its merger history, and has important implications for any dynamical modelling attempt of both its halo and disc populations. It also has important implications for dark matter direct detection experiments \cite{Dodge2023,Freese2013,Read2014}. Perspectives of measuring the tilting rate in external galaxies further open the possibility of constraining their merger history even when other signatures of past mergers are not visible anymore!

\section{Methods}\label{sec:methods}

\subsection{Gaia data selection}

The proper motions analysed in this work are based on the Gaia Data Release 3 (Gaia DR3) \cite{Gaia2023}. Starting from the 1.8 billion sources with astrometric measurements, we retained only objects with a five- or six-parameter astrometric solution (\texttt{astrometric\_params\_solved} $\geq 31$), high-quality astrometry (RUWE $< 1.4$), and no duplicated source flag. To minimise contamination from non-stellar or poorly resolved sources, we further excluded objects with evidence of complex image profiles (\texttt{ipd\_multi\_peak} $\leq 2$ and \texttt{ipd\_gof\_harmonic\_amplitude} $< 0.2$), as well as sources classified as candidate quasars or galaxies in the Gaia catalogue.

Heliocentric distances were estimated by inverting the parallaxes after correcting for the global parallax zero-point offset of $-0.017$~mas \cite{Lindegren2021}. To ensure reliable distance estimates, we restricted the sample to stars with \texttt{parallax\_over\_error} higher than 5. The final catalogue contains approximately 180 million stars spanning heliocentric distances up to 8~kpc.

We divided the sky into HEALPix \cite{Gorski2005} of level 5 in Galactic coordinates, each of them being subdivided into 1~kpc-wide heliocentric distance intervals between 1 and 8~kpc. The mean proper motion $\mu_b$ was then calculated independently in each pixel. The distance bin between 0 and 1~kpc was excluded from the analysis because the mean proper motion in Galactic latitude exhibits anomalous values, particularly towards the Galactic centre. This behaviour is likely related to residual systematic uncertainties in the parallax zero-point correction. As the stars in this distance range are very nearby, small systematic errors in the parallaxes translate into comparatively larger biases in the inferred distances and, consequently, in the measured mean proper motions.

\subsection{Milky Way models: GUMS and BV23}

To compare the observed mean proper motions with theoretical expectations, we considered two independent self-consistent models of the Milky Way.

The first model is the 20th version of the Gaia Universe Model Snapshot (GUMS) \cite{Robin2012}, obtained from the Gaia Archive. The second model, BV23 \cite{Binney2023}, was constructed with \texttt{AGAMA} \cite{Vasiliev2019} using one billion stellar particles. The model is constrained primarily by the kinematics of stars observed with the Gaia Radial Velocity Spectrometer (RVS) together with measurements of the stellar density in the Solar neighbourhood. Unlike GUMS, BV23 does not include a Galactic warp in its stellar distribution.

Note that GUMS provides the catalogue of simulated stars directly in observable coordinates. Therefore, to have a fair comparison between the two models we used, based on the same Solar location and velocity, we first recomputed the Galactocentric positions and velocities using the Solar reference frame described in the Gaia DR3 documentation.

For both models, Galactocentric coordinates were transformed into Galactic observables assuming a Solar Galactocentric distance of $R_\odot = 8.27$~kpc \cite{Gravity2021}, a Solar height above the Galactic mid-plane of $Z_\odot = 20$~pc \cite{Bennett2019}, and a Solar Galactocentric velocity of $(U_\odot, V_\odot, W_\odot) = (11.1, 251.0, 8.59)$~\kms. Note that $W_\odot$ was obtained by assuming that Sgr~A$^*$ is stationary at the Galactic centre and adopting the latest proper motion measurements from the literature \cite{Reid2020}. Finally, the mean proper motion $\mu_b$ of the simulated stars in GUMS and BV23 is computed as for the Gaia data.

\subsection{Inference of the tilting rate with a Monte Carlo Markov Chain}

The angular tilting rate was inferred using a Markov chain Monte Carlo (MCMC) analysis implemented with the \texttt{emcee} Python package \cite{ForemanMackey2019}. For each fit, the sampler was evolved for 2,500 steps, of which the first 500 were discarded as burn-in, leaving 2,000 production samples for the posterior analysis. The likelihood function was assumed to be Gaussian and is given by
\begin{equation*}
\log(\mathcal{L}) = -0.5 \sum_i \left( \langle \mu_{b,\mathrm{obs}} \rangle_i - \left( \langle \mu_{b,\mathrm{sim}} \rangle_i + \mu_{b,\mathrm{tilt},i} \right) \right)^2
\end{equation*}
where $\langle \mu_{b,\mathrm{obs}} \rangle$ and $\langle \mu_{b,\mathrm{sim}} \rangle$ denote the mean observed and simulated proper motions in Galactic latitude, respectively, and $\mu_{b,\mathrm{tilt}}$ is the contribution arising from the disc tilting, expressed as
\begin{equation*}
\mu_{b,\mathrm{tilt}}(l, b, d) = \left( \Omega_x A_x + \Omega_y A_y \right) / \kappa,
\end{equation*}
with $A_x = \sin(l)$ and $A_y = -\cos(l) + (R_\odot/d)\cos(b)$, where $l$ and $b$ are the Galactic longitude and latitude, $d$ is the heliocentric distance, $R_\odot$ is the Solar Galactocentric radius, and $\kappa = 4.74047$ converts an angular velocity from \kmskpc\ to mas\,yr$^{-1}$.

Uniform priors spanning $\pm 10$~\kmskpc\ were adopted for both $\Omega_x$ and $\Omega_y$. Only HEALPix pixels containing at least 10 stars observed by Gaia and at least three simulated particles were included in the fit, leading to between 1040 and 6016 fitted $(l,b)$ sky positions per distance bin. To assess the robustness of the inferred tilting rate, the analysis was repeated using several maximum Galactic latitude cuts, as shown in Fig.~\ref{fig:2}. We did not probe Galactic latitude above $|b| > 30^\circ$, as stellar halo contamination becomes increasingly important at higher latitudes, particularly beyond $d > 4$~kpc.

Allowing $\Omega_y$ to vary freely or fixing it to zero produced virtually identical estimates of $\Omega_x$ (with differences below 0.5\%). Throughout this work, quoted parameter values correspond to the median of the posterior distribution, while uncertainties denote the 16$^{\rm th}$ and 84$^{\rm th}$ percentiles.

\subsection{Comparison to numerical simulations}

To interpret the observed tilting rate of the Milky Way, we compared our measurements with a suite of collisionless $N$-body simulations of mergers between a Milky Way-like galaxy and a massive satellite similar to the progenitor of Gaia--Enceladus/Sausage. We explored three merger mass ratios: 1:2.5, 1:4 and 1:10. In all simulations, the initial Milky Way model was identical and based on an existing model \cite{Naidu2021}. For the 1:2.5 merger, the satellite progenitor is similar to the one used in this previous work, and includes a stellar disc. For the lower-mass mergers, however, the stellar disc was removed because its contribution to the disc tilting is negligible compared with that of the satellite's dark matter halo. The dark matter halo concentration was adjusted for each progenitor according to an established mass--concentration relation \cite{MunozCuartas2011}. The adopted structural parameters are listed in Extended Data Table~\ref{tab:ED1}. Both galaxies were constructed as fully self-consistent equilibrium models using the \texttt{AGAMA} dynamical framework \cite{Vasiliev2019}, with the individual velocities of the bulge and dark matter haloes obtained using the quasi-spherical distribution function, while the quasi-isothermal distribution function was used for the stellar discs, with a radial velocity-dispersion scale length of $R_{\sigma,r} = 2R_\mathrm{disc}$, and a central radial velocity dispersion of $\sigma_{r,0}$ listed in Extended Data Table~\ref{tab:ED1}. All stellar and dark matter particles were assigned an identical mass of $10^{4.5}$~M$_\odot$, and each galaxy was evolved in isolation for 3~Gyr to verify its dynamical stability before the merger simulations.

The satellite was initially placed at the virial radius of the Milky Way halo with a total velocity equal to the circular velocity at this distance ($r_{200}$). Following previous work \cite{Naidu2021}, its orbit had a circularity, corresponding to the ratio of the tangential velocity to the total velocity, of 0.5, starting with an inclination of $15^\circ$ from the Galactic disc. We explored both prograde and retrograde encounters over 20 inclinations of the orbital angular momentum ($\theta_\mathrm{orb} = 0^\circ$, $\pm 30^\circ$, $\pm 60^\circ$, $\pm 75^\circ$, $\pm 83^\circ$, $\pm 90^\circ$, $\pm 97^\circ$, $\pm 105^\circ$, $\pm 120^\circ$, $\pm 150^\circ$, $180^\circ$), allowing us to sample the variety of merger geometries expected in a cosmological context. The simulations were run with \texttt{gyrfalcON} \cite{Dehnen2000,Dehnen2002}, included in the \texttt{NEMO} framework, over 10~Gyr. Individual particle timesteps were assigned adaptively according to the local gravitational acceleration with a maximum timestep of 15.625~Myr. Gravitational forces were softened using the recommended P1 kernel with a softening length of 250~pc (equivalent to a Plummer softening length of 177~pc).

At each simulation snapshot, the system was recentered on the Milky Way using the mean position and velocity of the bulge particles. The orientation of the Galactic disc, characterised by the longitude, $\theta_\mathrm{disc}$, and colatitude, $\varphi_\mathrm{disc}$, was then determined iteratively using a principal component analysis (PCA) of the stellar disc. The first iteration included all disc particles within 10~kpc of the Galactic centre. Subsequent iterations retained only particles with a prograde vertical angular momentum $L_z > 500$~km\,s$^{-1}$\,kpc, thereby excluding stars that had been strongly perturbed by the merger \cite{Villalobos2008}. The procedure was repeated until the change in the disc orientation between successive iterations was smaller than $0.5^\circ$. The evolution of the angular momentum direction of the disc is shown in the Extended Data Fig.~\ref{fig:ED1}. Note that in some conditions, in particular when the Galactic disc is highly inclined compared to the orbital direction of the satellite, the disc can flip by more than $90^\circ$ \cite{Batrakov2026}.

The time derivatives of both angles were then estimated using a Savitzky--Golay filter with a quadratic polynomial and a window of 9 snapshots (corresponding to $\sim 0.5$~Gyr), which simultaneously smooths numerical fluctuations and computes the derivatives. The instantaneous tilting rate of the disc, $\Omega_\mathrm{disc}$, was finally obtained from the spherical metric,
\begin{equation*}
\Omega_\mathrm{disc} = \sqrt{\dot{\varphi}^2 + \dot{\theta}^2 \sin^2(\varphi)},
\end{equation*}
which gives the true angular velocity of the disc's orientation on the unit sphere. With this method, an isolated Milky Way model, evolved for 3~Gyr, tumbles around a stable orientation with an angular velocity of 1.3\degGyr, which can be considered the systematic error of our method.

\section{Extended data}\label{sec:ED}

\setcounter{table}{0}
\setcounter{figure}{0}
\renewcommand{\tablename}{Extended Data Table}
\renewcommand{\figurename}{Extended Data Fig.}
\renewcommand{\thetable}{\arabic{table}}
\renewcommand{\thefigure}{\arabic{figure}}

\begin{table}[htbp]
\caption{\textbf{Initial parameters of the simulated galaxies.}}\label{tab:ED1}
\footnotesize
\begin{tabular*}{\textwidth}{@{\extracolsep\fill}llllllll}
\toprule
\multicolumn{3}{@{}c}{Milky Way} & \multicolumn{5}{c@{}}{Gaia-Enceladus/Sausage} \\
\cmidrule{1-3}\cmidrule{4-8}
Component & Parameter & Value & Component & Parameter & 1:2.5 & 1:4 & 1:10 \\
\midrule
DM halo     &                                   &       & DM halo     &                                   &       &       &       \\
Hernquist   & Mass [$10^{11}$ M$_\odot$]        & 5.0   & Hernquist   & Mass [$10^{11}$ M$_\odot$]        & 2.0   & 1.25  & 0.5   \\
            & Scale length ($a$) [kpc]          & 42.3  &             & Scale length ($a$) [kpc]          & 30.25 & 23.57 & 16.98 \\
            & Concentration ($c_{200}$)         & 3.8   &             & Concentration ($c_{200}$)         & 4.0   & 4.65  & 4.82  \\
\addlinespace
Disc        &                                   &       & Disc        &                                   & Yes   & No    & No    \\
Exponential & Mass [$10^{9}$ M$_\odot$]         & 6.0   & Exponential & Mass [$10^{8}$ M$_\odot$]         & 5.0   &       &       \\
            & $R_\mathrm{disc}$ [kpc]           & 2.0   &             & $R_\mathrm{disc}$ [kpc]           & 1.7   &       &       \\
            & $h_\mathrm{disc}$ [kpc]           & 1.0   &             & $h_\mathrm{disc}$ [kpc]           & 01.02 &       &       \\
            & $\sigma_{r,0}$ [\kms]             & 100.0 &             & $\sigma_{r,0}$ [\kms]             & 50.0  &       &       \\
\addlinespace
Bulge       &                                   &       &             &                                   &       &       &       \\
Hernquist   & Mass [$10^{10}$ M$_\odot$]        & 1.4   &             &                                   &       &       &       \\
            & Scale length ($a$) [kpc]          & 1.5   &             &                                   &       &       &       \\
\botrule
\end{tabular*}
\end{table}

\begin{figure}[htbp]
\centering
\includegraphics[width=\textwidth]{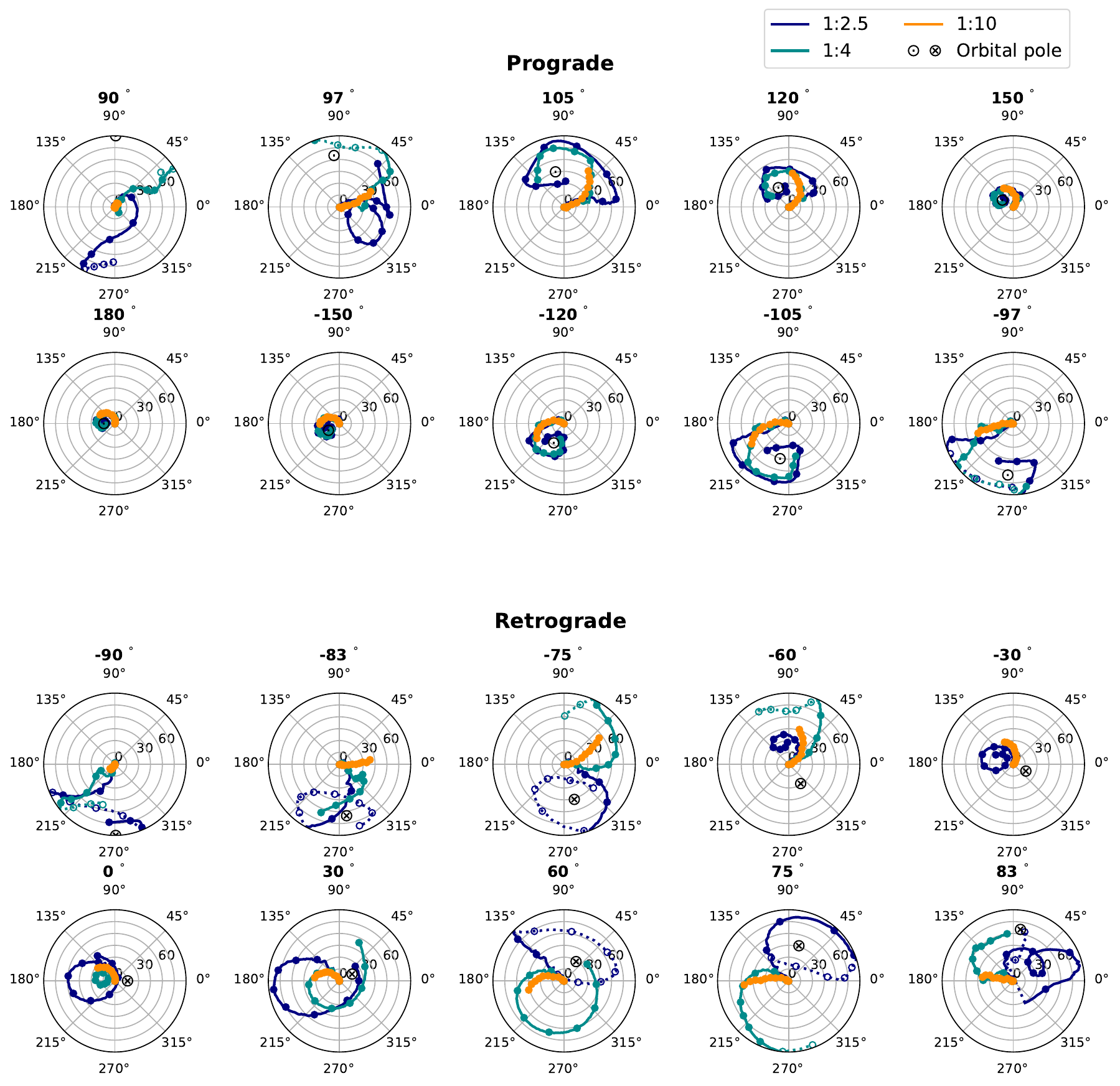}
\caption{\textbf{Evolution of the Milky Way disc's angular momentum direction following a satellite merger.} Each polar panel shows the trajectory of the disc's angular momentum direction ($\theta_\mathrm{disc}$ and $\varphi_\mathrm{disc}$) as a function of time, for a satellite accreted on a prograde (top block) or retrograde (bottom block) orbit with the orbital inclination $\theta_\mathrm{orb}$ indicated above each panel. Filled circles mark the disc orientation at 1~Gyr intervals; open circles indicate the same when the trajectory has crossed into the opposite hemisphere of pitch. Solid lines trace the disc orientation while its precession maintains its original sense; dashed lines indicate the trajectory after the orientation has flipped to the opposite sign, i.e. once the disc has precessed past the pole. The $\odot$ (prograde) and $\otimes$ (retrograde) symbols mark the initial direction of the orbital angular momentum of the accreted satellite.}\label{fig:ED1}
\end{figure}

\section{Supplementary material}\label{sec:SI}

% Supplementary figures are numbered S1, S2, ...
\setcounter{figure}{0}
\renewcommand{\figurename}{Fig.}
\renewcommand{\thefigure}{S\arabic{figure}}

\subsection{Tentative measurement of $\Omega_y$}

\begin{figure}[htbp]
\centering
\includegraphics[width=0.9\textwidth]{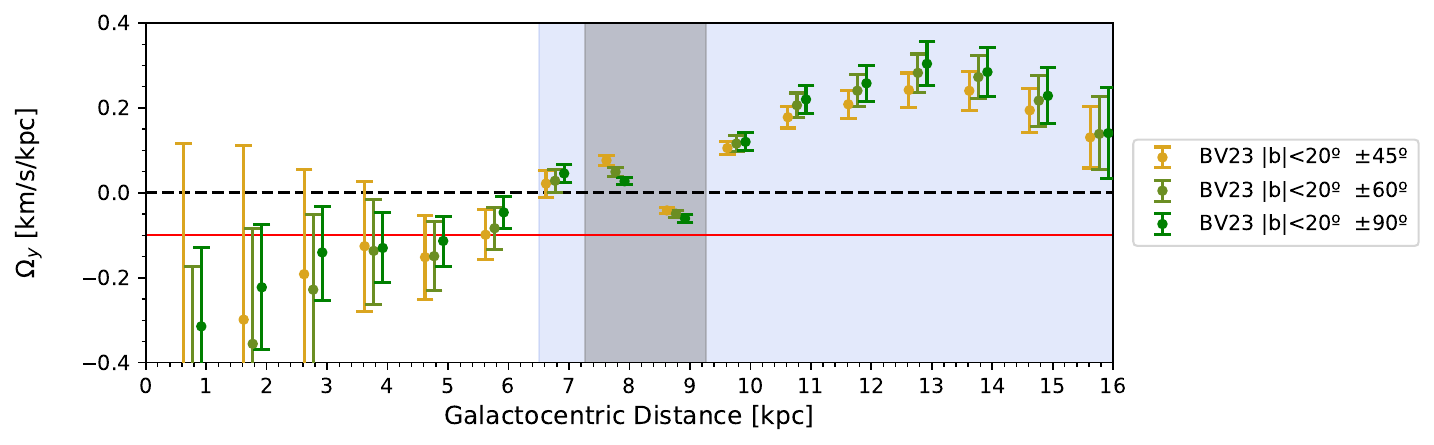}
\caption{\textbf{Present-day tilting rate of the Milky Way disc about the Y-axis.} Measurements of $\Omega_y$ obtained using the methodology described in the Methods section, shown as a function of Galactocentric radius, $R$. Results are presented separately for three cones of $\pm 45^\circ$, $60^\circ$ and $90^\circ$ towards the Galactic centre and anticentre, as indicated in the legend. Points represent the median of the posterior distribution, while error bars denote the 16th and 84th percentiles. The grey shaded region marks the 1~kpc-wide interval around the Solar radius, where the measurements are considered unreliable. The blue shaded region indicates $R > 6.5$~kpc, where the vertical kinematic signature of the Galactic warp becomes significant \cite{Nibauer2024,Perryman2014}. The horizontal dotted line corresponds to the null hypothesis of no disc tilting ($\Omega_y = 0$~\kmskpc), whereas the red line indicates the tentative value measured with our method of $\Omega_y = -0.1$~\kmskpc.}\label{fig:S1}
\end{figure}

As discussed in the main text, measuring $\Omega_y$ is particularly challenging because its kinematic signature is strongly contaminated by the Galactic warp, whose amplitude is similar, if not larger than, the expected tilting rate. This is illustrated in Fig.~\ref{fig:S1}, where $\Omega_y$ is measured at different Galactocentric radii using the same methodology as described for $\Omega_x$. Beyond $R > 6.5$~kpc (blue shaded region), the kinematic signature of the Galactic warp becomes clearly apparent as a wave-like variation with radius, consistent with previous studies \cite{Schonrich2018,Huang2018}.

At smaller Galactocentric radii, the inferred values of $\Omega_y$ appear to approach a plateau of $\Omega_y \simeq -0.1$~\kmskpc\ (6\degGyr), although with substantial uncertainties. We interpret this plateau as a tentative upper limit on the absolute amplitude of $\Omega_y$ rather than as a robust detection. In any case, our results indicate that the present-day tilting of the Galactic disc is dominated by rotation about the X-axis, with any contribution from $\Omega_y$ remaining small in comparison.

\backmatter

\bmhead{Acknowledgements}

The authors thank Andrea Negri and Claudio Della Vecchia for their assistance with the $N$-body simulations.

\bmhead{Funding statement}

GFT acknowledges the Ram\'on y Cajal programme (grant RYC2024-051016-I), funded by MICIU/AEI/10.13039/501100011033 and by the European Social Fund Plus (ESF+). GFT, AR and GB acknowledge support from the Agencia Estatal de Investigac\'ion del Ministerio de Ciencia e Innovac\'ion (AEI-MCIN) under grant number PID2023-150319NB-C21 and PID2023-150319NB-C22. This project has received funding from the European Union's Horizon Europe research and innovation programme under Grant Agreement No. 101158446 (ExGal-Twin). This work is part of grant CEX2025-001609-S, awarded to the Instituto de Astrof\'isica de Canarias under the Severo Ochoa Centre of Excellence program and funded by MICIU/AEI/10.13039/501100011033.

\bmhead{Author contributions}

All authors assisted in the development and writing of the paper. In addition, the observational selection and the method were developed by G.F.T, B.F., and R.~I., the $N$-body simulations were run by G.F.T., with the assistance of A.R., and analysed with G.B.

\bmhead{Competing Interests}

The authors have no competing financial interests.

\clearpage
\bibliography{sn-bibliography}

@ARTICLE{Eggen1962,
       author = {{Eggen}, O.~J. and {Lynden-Bell}, D. and {Sandage}, A.~R.},
        title = "{Evidence from the motions of old stars that the Galaxy collapsed.}",
      journal = {Astrophys. J.},
         year = 1962,
        month = nov,
       volume = {136},
        pages = {748},
          doi = {10.1086/147433},
       adsurl = {https://ui.adsabs.harvard.edu/abs/1962ApJ...136..748E}
}

@ARTICLE{Searle1978,
       author = {{Searle}, L. and {Zinn}, R.},
        title = "{Composition of halo clusters and the formation of the galactic halo.}",
      journal = {Astrophys. J.},
         year = 1978,
        month = oct,
       volume = {225},
        pages = {357-379},
          doi = {10.1086/156499},
       adsurl = {https://ui.adsabs.harvard.edu/abs/1978ApJ...225..357S}
}

@ARTICLE{White1978,
       author = {{White}, S.~D.~M. and {Rees}, M.~J.},
        title = "{Core condensation in heavy halos: a two-stage theory for galaxy formation and clustering.}",
      journal = {Mon. Not. R. Astron. Soc.},
         year = 1978,
        month = may,
       volume = {183},
        pages = {341-358},
          doi = {10.1093/mnras/183.3.341},
       adsurl = {https://ui.adsabs.harvard.edu/abs/1978MNRAS.183..341W}
}

@ARTICLE{White1991,
       author = {{White}, Simon D.~M. and {Frenk}, Carlos S.},
        title = "{Galaxy Formation through Hierarchical Clustering}",
      journal = {Astrophys. J.},
         year = 1991,
        month = sep,
       volume = {379},
        pages = {52},
          doi = {10.1086/170483},
       adsurl = {https://ui.adsabs.harvard.edu/abs/1991ApJ...379...52W}
}

@ARTICLE{Springel2005,
       author = {{Springel}, Volker and {Hernquist}, Lars},
        title = "{Formation of a Spiral Galaxy in a Major Merger}",
      journal = {Astrophys. J. Lett.},
         year = 2005,
        month = mar,
       volume = {622},
       number = {1},
        pages = {L9-L12},
          doi = {10.1086/429486},
archivePrefix = {arXiv},
       eprint = {astro-ph/0411379},
 primaryClass = {astro-ph},
       adsurl = {https://ui.adsabs.harvard.edu/abs/2005ApJ...622L...9S}
}

@article{Bullock2005,
  author  = {Bullock, J. S. and Johnston, K. V.},
  title   = {Tracing galaxy formation with stellar halos. {I}. {M}ethods},
  journal = {Astrophys. J.},
  volume  = {635},
  pages   = {931--949},
  year    = {2005}
}

@article{Gomez2017,
  author  = {{G{\'o}mez}, Facundo A. and {White}, Simon D.~M. and {Grand}, Robert J.~J. and {Marinacci}, Federico and {Springel}, Volker and {Pakmor}, R{\"u}diger},
  title   = {Warps and waves in the stellar discs of the {A}uriga cosmological simulations},
  journal = {Mon. Not. R. Astron. Soc.},
  volume  = {465},
  pages   = {3446--3460},
  year    = {2017}
}

@article{Antoja2018,
  author  = {{Antoja}, T. and {Helmi}, A. and {Romero-G{\'o}mez}, M. and {Katz}, D. and {Babusiaux}, C. and {Drimmel}, R. and {Evans}, D.~W. and {Figueras}, F. and {Poggio}, E. and {Reyl{\'e}}, C. and {Robin}, A.~C. and {Seabroke}, G. and {Soubiran}, C.},
  title   = {A dynamically young and perturbed {M}ilky {W}ay disk},
  journal = {Nature},
  volume  = {561},
  pages   = {360--362},
  year    = {2018}
}

@article{Laporte2019,
  author  = {Laporte, C. F. P. and Johnston, K. V. and Tzanidakis, A.},
  title   = {Stellar disc streams as probes of the {G}alactic potential and satellite impacts},
  journal = {Mon. Not. R. Astron. Soc.},
  volume  = {483},
  pages   = {1427--1436},
  year    = {2019}
}

@article{Poggio2020,
  author  = {{Poggio}, E. and {Drimmel}, R. and {Andrae}, R. and {Bailer-Jones}, C.~A.~L. and {Fouesneau}, M. and {Lattanzi}, M.~G. and {Smart}, R.~L. and {Spagna}, A.},
  title   = {Evidence of a dynamically evolving {G}alactic warp},
  journal = {Nat. Astron.},
  volume  = {4},
  pages   = {590--596},
  year    = {2020}
}

@article{Poggio2025,
  author  = {{Poggio}, E. and {Khanna}, S. and {Drimmel}, R. and {Zari}, E. and {D'Onghia}, E. and {Lattanzi}, M.~G. and {Palicio}, P.~A. and {Recio-Blanco}, A. and {Thulasidharan}, L.},
  title   = {The great wave: evidence of a large-scale vertical corrugation propagating outwards in the {G}alactic disc},
  journal = {Astron. Astrophys.},
  volume  = {699},
  pages   = {A199},
  year    = {2025}
}

@article{Binney1986,
  author  = {Binney, J. and May, A.},
  title   = {The spheroids of galaxies before and after disc formation},
  journal = {Mon. Not. R. Astron. Soc.},
  volume  = {218},
  pages   = {743--760},
  year    = {1986}
}

@article{Ostriker1989,
  author  = {Ostriker, E. C. and Binney, J. J.},
  title   = {Warped and tilted galactic discs},
  journal = {Mon. Not. R. Astron. Soc.},
  volume  = {237},
  pages   = {785--798},
  year    = {1989}
}

@article{Huang1997,
  author  = {Huang, S. and Carlberg, R. G.},
  title   = {Sinking satellites and tilting disk galaxies},
  journal = {Astrophys. J.},
  volume  = {480},
  pages   = {503--523},
  year    = {1997}
}

@article{Sellwood1998,
  author  = {Sellwood, J. A. and Nelson, R. W. and Tremaine, S.},
  title   = {Resonant thickening of disks by small satellite galaxies},
  journal = {Astrophys. J.},
  volume  = {506},
  pages   = {590--599},
  year    = {1998}
}

@article{Dodge2023,
  author  = {Dodge, B. C. and Slone, O. and Lisanti, M. and Cohen, T.},
  title   = {Dynamics of stellar disc tilting from satellite mergers},
  journal = {Mon. Not. R. Astron. Soc.},
  volume  = {518},
  pages   = {2870--2884},
  year    = {2023}
}

@article{Gaia2021,
  author  = {{Gaia Collaboration} and {Brown}, A.~G.~A. and {Vallenari}, A. and {Prusti}, T. and {de Bruijne}, J.~H.~J. and {Babusiaux}, C. and {Biermann}, M. and {Creevey}, O.~L. and {Evans}, D.~W. and {Eyer}, L. and {Hutton}, A. and {Jansen}, F. and {Jordi}, C. and {Klioner}, S.~A. and {Lammers}, U. and {Lindegren}, L. and {Luri}, X. and {Mignard}, F. and {Panem}, C. and {Pourbaix}, D. and {Randich}, S. and {Sartoretti}, P. and {Soubiran}, C. and {Walton}, N.~A. and {Arenou}, F. and {Bailer-Jones}, C.~A.~L. and {Bastian}, U. and {Cropper}, M. and {Drimmel}, R. and {Katz}, D. and {Lattanzi}, M.~G. and {van Leeuwen}, F. and {Bakker}, J. and {Cacciari}, C. and {Casta{\~n}eda}, J. and {De Angeli}, F. and {Ducourant}, C. and {Fabricius}, C. and {Fouesneau}, M. and {Fr{\'e}mat}, Y. and {Guerra}, R. and {Guerrier}, A. and {Guiraud}, J. and {Jean-Antoine Piccolo}, A. and {Masana}, E. and {Messineo}, R. and {Mowlavi}, N. and {Nicolas}, C. and {Nienartowicz}, K. and {Pailler}, F. and {Panuzzo}, P. and {Riclet}, F. and {Roux}, W. and {Seabroke}, G.~M. and {Sordo}, R. and {Tanga}, P. and {Th{\'e}venin}, F. and {Gracia-Abril}, G. and {Portell}, J. and {Teyssier}, D. and {Altmann}, M. and {Andrae}, R. and {Bellas-Velidis}, I. and {Benson}, K. and {Berthier}, J. and {Blomme}, R. and {Brugaletta}, E. and {Burgess}, P.~W. and {Busso}, G. and {Carry}, B. and {Cellino}, A. and {Cheek}, N. and {Clementini}, G. and {Damerdji}, Y. and {Davidson}, M. and {Delchambre}, L. and {Dell'Oro}, A. and {Fern{\'a}ndez-Hern{\'a}ndez}, J. and {Galluccio}, L. and {Garc{\'\i}a-Lario}, P. and {Garcia-Reinaldos}, M. and {Gonz{\'a}lez-N{\'u}{\~n}ez}, J. and {Gosset}, E. and {Haigron}, R. and {Halbwachs}, J.-L. and {Hambly}, N.~C. and {Harrison}, D.~L. and {Hatzidimitriou}, D. and {Heiter}, U. and {Hern{\'a}ndez}, J. and {Hestroffer}, D. and {Hodgkin}, S.~T. and {Holl}, B. and {Jan{\ss}en}, K. and {Jevardat de Fombelle}, G. and {Jordan}, S. and {Krone-Martins}, A. and {Lanzafame}, A.~C. and {L{\"o}ffler}, W. and {Lorca}, A. and {Manteiga}, M. and {Marchal}, O. and {Marrese}, P.~M. and {Moitinho}, A. and {Mora}, A. and {Muinonen}, K. and {Osborne}, P. and {Pancino}, E. and {Pauwels}, T. and {Petit}, J.-M. and {Recio-Blanco}, A. and {Richards}, P.~J. and {Riello}, M. and {Rimoldini}, L. and {Robin}, A.~C. and {Roegiers}, T. and {Rybizki}, J. and {Sarro}, L.~M. and {Siopis}, C. and {Smith}, M. and {Sozzetti}, A. and {Ulla}, A. and {Utrilla}, E. and {van Leeuwen}, M. and {van Reeven}, W. and {Abbas}, U. and {Abreu Aramburu}, A. and {Accart}, S. and {Aerts}, C. and {Aguado}, J.~J. and {Ajaj}, M. and {Altavilla}, G. and {{\'A}lvarez}, M.~A. and {{\'A}lvarez Cid-Fuentes}, J. and {Alves}, J. and {Anderson}, R.~I. and {Anglada Varela}, E. and {Antoja}, T. and {Audard}, M. and {Baines}, D. and {Baker}, S.~G. and {Balaguer-N{\'u}{\~n}ez}, L. and {Balbinot}, E. and {Balog}, Z. and {Barache}, C. and {Barbato}, D. and {Barros}, M. and {Barstow}, M.~A. and {Bartolom{\'e}}, S. and {Bassilana}, J.-L. and {Bauchet}, N. and {Baudesson-Stella}, A. and {Becciani}, U. and {Bellazzini}, M. and {Bernet}, M. and {Bertone}, S. and {Bianchi}, L. and {Blanco-Cuaresma}, S. and {Boch}, T. and {Bombrun}, A. and {Bossini}, D. and {Bouquillon}, S. and {Bragaglia}, A. and {Bramante}, L. and {Breedt}, E. and {Bressan}, A. and {Brouillet}, N. and {Bucciarelli}, B. and {Burlacu}, A. and {Busonero}, D. and {Butkevich}, A.~G. and {Buzzi}, R. and {Caffau}, E. and {Cancelliere}, R. and {C{\'a}novas}, H. and {Cantat-Gaudin}, T. and {Carballo}, R. and {Carlucci}, T. and {Carnerero}, M.~I. and {Carrasco}, J.~M. and {Casamiquela}, L. and {Castellani}, M. and {Castro-Ginard}, A. and {Castro Sampol}, P. and {Chaoul}, L. and {Charlot}, P. and {Chemin}, L. and {Chiavassa}, A. and {Cioni}, M.-R.~L. and {Comoretto}, G. and {Cooper}, W.~J. and {Cornez}, T. and {Cowell}, S. and {Crifo}, F. and {Crosta}, M. and {Crowley}, C. and {Dafonte}, C. and {Dapergolas}, A. and {David}, M. and {David}, P.},
  title   = {{Gaia} {E}arly {D}ata {R}elease 3. {S}ummary of the contents and survey properties},
  journal = {Astron. Astrophys.},
  volume  = {649},
  pages   = {A1},
  year    = {2021}
}

@article{Earp2017,
  author  = {Earp, S. W. F. and Debattista, V. P. and Maccio, A. V. and Cole, D. R.},
  title   = {The tilting rate of the {M}ilky {W}ay's disc},
  journal = {Mon. Not. R. Astron. Soc.},
  volume  = {469},
  pages   = {4095--4101},
  year    = {2017}
}

@article{Dillamore2022,
  author  = {Dillamore, A. M. and Belokurov, V. and Font, A. S. and McCarthy, I. G.},
  title   = {Merger-induced galaxy transformations in the {ARTEMIS} simulations},
  journal = {Mon. Not. R. Astron. Soc.},
  volume  = {513},
  pages   = {1867--1886},
  year    = {2022}
}

@article{Belokurov2018,
  author  = {Belokurov, V. and Erkal, D. and Evans, N. W. and Koposov, S. E. and Deason, A. J.},
  title   = {Co-formation of the disc and the stellar halo},
  journal = {Mon. Not. R. Astron. Soc.},
  volume  = {478},
  pages   = {611--619},
  year    = {2018}
}

@article{Helmi2018,
  author  = {{Helmi}, Amina and {Babusiaux}, Carine and {Koppelman}, Helmer H. and {Massari}, Davide and {Veljanoski}, Jovan and {Brown}, Anthony G.~A.},
  title   = {The merger that led to the formation of the {M}ilky {W}ay's inner stellar halo and thick disk},
  journal = {Nature},
  volume  = {563},
  pages   = {85--88},
  year    = {2018}
}

@article{Gallart2019,
  author  = {{Gallart}, Carme and {Bernard}, Edouard J. and {Brook}, Chris B. and {Ruiz-Lara}, Tom{\'a}s and {Cassisi}, Santi and {Hill}, Vanessa and {Monelli}, Matteo},
  title   = {Uncovering the birth of the {M}ilky {W}ay through accurate stellar ages with {Gaia}},
  journal = {Nat. Astron.},
  volume  = {3},
  pages   = {932--939},
  year    = {2019}
}

@article{DiMatteo2019,
  author  = {{Di Matteo}, P. and {Haywood}, M. and {Lehnert}, M.~D. and {Katz}, D. and {Khoperskov}, S. and {Snaith}, O.~N. and {G{\'o}mez}, A. and {Robichon}, N.},
  title   = {The {M}ilky {W}ay has no in-situ halo other than the heated thick disc},
  journal = {Astron. Astrophys.},
  volume  = {632},
  pages   = {A4},
  year    = {2019}
}

@article{Wang2026,
  author  =  {{Wang}, Yuan and {Luo}, Xiong and {Wang}, Huiyuan and {Wang}, Enci and {Li}, Hao and {Marinacci}, Federico and {Shen}, Xuejian and {Vogelsberger}, Mark},
  title   = {A universal dance of galactic disks: ubiquitous precession and its implications},
  journal = {Astrophys. J.},
  volume  = {1006},
  pages   = {160},
  year    = {2026}
}

@article{Nibauer2024,
  author  = {Nibauer, J. and Bonaca, A. and Lisanti, M. and Erkal, D. and Hastings, Z.},
  title   = {Slant, fan, and narrow: the response of stellar streams to a tilting {G}alactic disk},
  journal = {Astrophys. J.},
  volume  = {969},
  pages   = {55},
  year    = {2024}
}

@article{Perryman2014,
  author  = {Perryman, M. and Spergel, D. N. and Lindegren, L.},
  title   = {The {Gaia} inertial reference frame and the tilting of the {M}ilky {W}ay disk},
  journal = {Astrophys. J.},
  volume  = {789},
  pages   = {166},
  year    = {2014}
}

@article{Schonrich2018,
  author  = {Sch{\"o}nrich, R. and Dehnen, W.},
  title   = {Warp, waves, and wrinkles in the {M}ilky {W}ay},
  journal = {Mon. Not. R. Astron. Soc.},
  volume  = {478},
  pages   = {3809--3824},
  year    = {2018}
}

@article{Huang2018,
  author  = {{Huang}, Y. and {Sch{\"o}nrich}, R. and {Liu}, X.-W. and {Chen}, B.-Q. and {Zhang}, H.-W. and {Yuan}, H.-B. and {Xiang}, M.-S. and {Wang}, C. and {Tian}, Z.-J.},
  title   = {On the kinematic signature of the {G}alactic warp as revealed by the {LAMOST-TGAS} data},
  journal = {Astrophys. J.},
  volume  = {864},
  pages   = {129},
  year    = {2018}
}

@article{Hunt2025,
  author  = {Hunt, J. A. S. and Vasiliev, E.},
  title   = {{M}ilky {W}ay dynamics in light of {Gaia}},
  journal = {New Astron. Rev.},
  volume  = {100},
  pages   = {101721},
  year    = {2025}
}

@article{Drimmel2001,
  author  = {Drimmel, R. and Spergel, D. N.},
  title   = {Three-dimensional structure of the {M}ilky {W}ay disk: The distribution of stars and dust beyond 0.35 {$R_\odot$}},
  journal = {Astrophys. J.},
  volume  = {556},
  pages   = {181--202},
  year    = {2001}
}

@article{Reyle2009,
  author  = {Reyl{\'e}, C. and Marshall, D. J. and Robin, A. C. and Schultheis, M.},
  title   = {The {M}ilky {W}ay's external disc constrained by {2MASS} star counts},
  journal = {Astron. Astrophys.},
  volume  = {495},
  pages   = {819--826},
  year    = {2009}
}

@article{Amores2017,
  author  = {Am{\^o}res, E. B. and Robin, A. C. and Reyl{\'e}, C.},
  title   = {Evolution over time of the {M}ilky {W}ay's disc shape},
  journal = {Astron. Astrophys.},
  volume  = {602},
  pages   = {A67},
  year    = {2017}
}

@article{Cheng2020,
  author  = {{Cheng}, Xinlun and {Anguiano}, Borja and {Majewski}, Steven R. and {Hayes}, Christian and {Arras}, Phil and {Chiappini}, Cristina and {Hasselquist}, Sten and {de Andrade Queiroz}, Anna B{\'a}rbara and {Nitschelm}, Christian and {Garc{\'\i}a-Hern{\'a}ndez}, Domingo An{\'\i}bal and {Lane}, Richard R. and {Roman-Lopes}, Alexandre and {Frinchaboy}, Peter M.},
  title   = {Exploring the {G}alactic warp through asymmetries in the kinematics of the {G}alactic disk},
  journal = {Astrophys. J.},
  volume  = {905},
  pages   = {49},
  year    = {2020}
}

@article{Dehnen2023,
  author  = {Dehnen, W. and Semczuk, M. and Sch{\"o}nrich, R.},
  title   = {A twisted and precessing {C}epheid warp in the outer {M}ilky {W}ay disc},
  journal = {Mon. Not. R. Astron. Soc.},
  volume  = {523},
  pages   = {1556--1564},
  year    = {2023}
}

@article{Binney2023,
  author  = {Binney, J. and Vasiliev, E.},
  title   = {Self-consistent models of our {G}alaxy},
  journal = {Mon. Not. R. Astron. Soc.},
  volume  = {520},
  pages   = {1832--1847},
  year    = {2023}
}

@article{Robin2012,
  author  = {{Robin}, A.~C. and {Luri}, X. and {Reyl{\'e}}, C. and {Isasi}, Y. and {Grux}, E. and {Blanco-Cuaresma}, S. and {Arenou}, F. and {Babusiaux}, C. and {Belcheva}, M. and {Drimmel}, R. and {Jordi}, C. and {Krone-Martins}, A. and {Masana}, E. and {Mauduit}, J.~C. and {Mignard}, F. and {Mowlavi}, N. and {Rocca-Volmerange}, B. and {Sartoretti}, P. and {Slezak}, E. and {Sozzetti}, A.},
  title   = {{Gaia} {U}niverse {M}odel {S}napshot: a statistical analysis of the expected contents of the {Gaia} catalogue},
  journal = {Astron. Astrophys.},
  volume  = {543},
  pages   = {A100},
  year    = {2012}
}

@article{Poggio2018,
  author  = {{Poggio}, E. and {Drimmel}, R. and {Lattanzi}, M.~G. and {Smart}, R.~L. and {Spagna}, A. and {Andrae}, R. and {Bailer-Jones}, C.~A.~L. and {Fouesneau}, M. and {Antoja}, T. and {Babusiaux}, C. and {Evans}, D.~W. and {Figueras}, F. and {Katz}, D. and {Reyl{\'e}}, C. and {Robin}, A.~C. and {Romero-G{\'o}mez}, M. and {Seabroke}, G.~M.},
  title   = {The {G}alactic warp revealed by {Gaia} {DR2} kinematics},
  journal = {Mon. Not. R. Astron. Soc.},
  volume  = {481},
  pages   = {L21--L25},
  year    = {2018}
}

@article{RomeroGomez2019,
  author  = {Romero-G{\'o}mez, M. and Mateu, C. and Aguilar, L. and Figueras, F. and Castro-Ginard, A.},
  title   = {{Gaia} kinematics reveal a complex lopsided and twisted {G}alactic disc warp},
  journal = {Astron. Astrophys.},
  volume  = {627},
  pages   = {A150},
  year    = {2019}
}

@article{CabreraGadea2024,
  author  = {Cabrera-Gadea, M. and Mateu, C. and Ramos, P. and Romero-G{\'o}mez, M. and Aguilar, L.},
  title   = {Structure, kinematics, and time evolution of the {G}alactic warp from {C}lassical {C}epheids},
  journal = {Mon. Not. R. Astron. Soc.},
  volume  = {528},
  pages   = {4409--4431},
  year    = {2024}
}

@article{Jonsson2024,
  author  = {J{\'o}nsson, V. H. and McMillan, P. J.},
  title   = {The tangled warp of the {M}ilky {W}ay},
  journal = {Astron. Astrophys.},
  volume  = {688},
  pages   = {A38},
  year    = {2024}
}

@article{Bett2012,
  author  = {Bett, P. E. and Frenk, C. S.},
  title   = {Spin flips---{I}. {E}volution of the angular momentum orientation of {M}ilky {W}ay-mass dark matter haloes},
  journal = {Mon. Not. R. Astron. Soc.},
  volume  = {420},
  pages   = {3324--3333},
  year    = {2012}
}

@article{Dekel2020,
  author  = {{Dekel}, Avishai and {Ginzburg}, Omri and {Jiang}, Fangzhou and {Freundlich}, Jonathan and {Lapiner}, Sharon and {Ceverino}, Daniel and {Primack}, Joel},
  title   = {A mass threshold for galactic gas discs by spin flips},
  journal = {Mon. Not. R. Astron. Soc.},
  volume  = {493},
  pages   = {4126--4142},
  year    = {2020}
}

@article{Zhu2026,
  author  = {{Zhu}, Ling and {Cai}, Runsheng and {Kang}, Xi and {Xue}, Xiang-Xiang and {Yang}, Chengqun and {Zhang}, Lan and {Mao}, Shude and {Liu}, Chao},
  title   = {A vertically orientated dark matter halo marks a flip of the {G}alactic disc},
  journal = {Astron. Astrophys.},
  volume  = {706},
  pages   = {A193},
  year    = {2026}
}

@article{Vasiliev2024,
  author  = {Vasiliev, E.},
  title   = {Dear {M}agellanic {C}louds, welcome back!},
  journal = {Mon. Not. R. Astron. Soc.},
  volume  = {527},
  pages   = {437--456},
  year    = {2024}
}

@article{Orkney2026,
  author  = {Orkney, M. D. A. and Laporte, C. F. P.},
  title   = {Build-up and survival of the disc: from numerical models of galaxy formation to the {M}ilky {W}ay},
  journal = {Mon. Not. R. Astron. Soc.},
  volume  = {548},
  pages   = {staf2154},
  year    = {2026}
}

@article{Freese2013,
  author  = {Freese, K. and Lisanti, M. and Savage, C.},
  title   = {Colloquium: {A}nnual modulation of dark matter},
  journal = {Rev. Mod. Phys.},
  volume  = {85},
  pages   = {1561--1581},
  year    = {2013}
}

@article{Read2014,
  author  = {Read, J. I.},
  title   = {The local dark matter density},
  journal = {J. Phys. G: Nucl. Part. Phys.},
  volume  = {41},
  pages   = {063101},
  year    = {2014}
}

@article{Gaia2023,
  author  =  {{Gaia Collaboration} and {Vallenari}, A. and {Brown}, A.~G.~A. and {Prusti}, T. and {de Bruijne}, J.~H.~J. and {Arenou}, F. and {Babusiaux}, C. and {Biermann}, M. and {Creevey}, O.~L. and {Ducourant}, C. and {Evans}, D.~W. and {Eyer}, L. and {Guerra}, R. and {Hutton}, A. and {Jordi}, C. and {Klioner}, S.~A. and {Lammers}, U.~L. and {Lindegren}, L. and {Luri}, X. and {Mignard}, F. and {Panem}, C. and {Pourbaix}, D. and {Randich}, S. and {Sartoretti}, P. and {Soubiran}, C. and {Tanga}, P. and {Walton}, N.~A. and {Bailer-Jones}, C.~A.~L. and {Bastian}, U. and {Drimmel}, R. and {Jansen}, F. and {Katz}, D. and {Lattanzi}, M.~G. and {van Leeuwen}, F. and {Bakker}, J. and {Cacciari}, C. and {Casta{\~n}eda}, J. and {De Angeli}, F. and {Fabricius}, C. and {Fouesneau}, M. and {Fr{\'e}mat}, Y. and {Galluccio}, L. and {Guerrier}, A. and {Heiter}, U. and {Masana}, E. and {Messineo}, R. and {Mowlavi}, N. and {Nicolas}, C. and {Nienartowicz}, K. and {Pailler}, F. and {Panuzzo}, P. and {Riclet}, F. and {Roux}, W. and {Seabroke}, G.~M. and {Sordo}, R. and {Th{\'e}venin}, F. and {Gracia-Abril}, G. and {Portell}, J. and {Teyssier}, D. and {Altmann}, M. and {Andrae}, R. and {Audard}, M. and {Bellas-Velidis}, I. and {Benson}, K. and {Berthier}, J. and {Blomme}, R. and {Burgess}, P.~W. and {Busonero}, D. and {Busso}, G. and {C{\'a}novas}, H. and {Carry}, B. and {Cellino}, A. and {Cheek}, N. and {Clementini}, G. and {Damerdji}, Y. and {Davidson}, M. and {de Teodoro}, P. and {Nu{\~n}ez Campos}, M. and {Delchambre}, L. and {Dell'Oro}, A. and {Esquej}, P. and {Fern{\'a}ndez-Hern{\'a}ndez}, J. and {Fraile}, E. and {Garabato}, D. and {Garc{\'\i}a-Lario}, P. and {Gosset}, E. and {Haigron}, R. and {Halbwachs}, J.-L. and {Hambly}, N.~C. and {Harrison}, D.~L. and {Hern{\'a}ndez}, J. and {Hestroffer}, D. and {Hodgkin}, S.~T. and {Holl}, B. and {Jan{\ss}en}, K. and {Jevardat de Fombelle}, G. and {Jordan}, S. and {Krone-Martins}, A. and {Lanzafame}, A.~C. and {L{\"o}ffler}, W. and {Marchal}, O. and {Marrese}, P.~M. and {Moitinho}, A. and {Muinonen}, K. and {Osborne}, P. and {Pancino}, E. and {Pauwels}, T. and {Recio-Blanco}, A. and {Reyl{\'e}}, C. and {Riello}, M. and {Rimoldini}, L. and {Roegiers}, T. and {Rybizki}, J. and {Sarro}, L.~M. and {Siopis}, C. and {Smith}, M. and {Sozzetti}, A. and {Utrilla}, E. and {van Leeuwen}, M. and {Abbas}, U. and {{\'A}brah{\'a}m}, P. and {Abreu Aramburu}, A. and {Aerts}, C. and {Aguado}, J.~J. and {Ajaj}, M. and {Aldea-Montero}, F. and {Altavilla}, G. and {{\'A}lvarez}, M.~A. and {Alves}, J. and {Anders}, F. and {Anderson}, R.~I. and {Anglada Varela}, E. and {Antoja}, T. and {Baines}, D. and {Baker}, S.~G. and {Balaguer-N{\'u}{\~n}ez}, L. and {Balbinot}, E. and {Balog}, Z. and {Barache}, C. and {Barbato}, D. and {Barros}, M. and {Barstow}, M.~A. and {Bartolom{\'e}}, S. and {Bassilana}, J.-L. and {Bauchet}, N. and {Becciani}, U. and {Bellazzini}, M. and {Berihuete}, A. and {Bernet}, M. and {Bertone}, S. and {Bianchi}, L. and {Binnenfeld}, A. and {Blanco-Cuaresma}, S. and {Blazere}, A. and {Boch}, T. and {Bombrun}, A. and {Bossini}, D. and {Bouquillon}, S. and {Bragaglia}, A. and {Bramante}, L. and {Breedt}, E. and {Bressan}, A. and {Brouillet}, N. and {Brugaletta}, E. and {Bucciarelli}, B. and {Burlacu}, A. and {Butkevich}, A.~G. and {Buzzi}, R. and {Caffau}, E. and {Cancelliere}, R. and {Cantat-Gaudin}, T. and {Carballo}, R. and {Carlucci}, T. and {Carnerero}, M.~I. and {Carrasco}, J.~M. and {Casamiquela}, L. and {Castellani}, M. and {Castro-Ginard}, A. and {Chaoul}, L. and {Charlot}, P. and {Chemin}, L. and {Chiaramida}, V. and {Chiavassa}, A. and {Chornay}, N. and {Comoretto}, G. and {Contursi}, G. and {Cooper}, W.~J. and {Cornez}, T. and {Cowell}, S. and {Crifo}, F. and {Cropper}, M. and {Crosta}, M. and {Crowley}, C. and {Dafonte}, C. and {Dapergolas}, A. and {David}, M. and {David}, P. and {de Laverny}, P. and {De Luise}, F. and {De March}, R.},
  title   = {{Gaia} {D}ata {R}elease 3: {S}ummary of the content and survey properties},
  journal = {Astron. Astrophys.},
  volume  = {674},
  pages   = {A1},
  year    = {2023}
}

@article{Lindegren2021,
  author  = {{Lindegren}, L. and {Bastian}, U. and {Biermann}, M. and {Bombrun}, A. and {de Torres}, A. and {Gerlach}, E. and {Geyer}, R. and {Hern{\'a}ndez}, J. and {Hilger}, T. and {Hobbs}, D. and {Klioner}, S.~A. and {Lammers}, U. and {McMillan}, P.~J. and {Ramos-Lerate}, M. and {Steidelm{\"u}ller}, H. and {Stephenson}, C.~A. and {van Leeuwen}, F.},
  title   = {{Gaia} {E}arly {D}ata {R}elease 3. {P}arallax bias versus magnitude, colour, and position},
  journal = {Astron. Astrophys.},
  volume  = {649},
  pages   = {A4},
  year    = {2021}
}

@article{Gorski2005,
  author  = {{G{\'o}rski}, K.~M. and {Hivon}, E. and {Banday}, A.~J. and {Wandelt}, B.~D. and {Hansen}, F.~K. and {Reinecke}, M. and {Bartelmann}, M.},
  title   = {{HEALPix}: A Framework for High-Resolution Discretization and Fast Analysis of Data Distributed on the Sphere},
  journal = {Astrophys. J.},
  volume  = {622},
  pages   = {2},
  year    = {2005}
}

@article{Vasiliev2019,
  author  = {Vasiliev, E.},
  title   = {{AGAMA}: action-based galaxy modelling architecture},
  journal = {Mon. Not. R. Astron. Soc.},
  volume  = {482},
  pages   = {1525--1544},
  year    = {2019}
}

@article{Gravity2021,
  author  =  {{GRAVITY Collaboration} and {Abuter}, R. and {Amorim}, A. and {Baub{\"o}ck}, M. and {Berger}, J.~P. and {Bonnet}, H. and {Brandner}, W. and {Cl{\'e}net}, Y. and {Davies}, R. and {de Zeeuw}, P.~T. and {Dexter}, J. and {Dallilar}, Y. and {Drescher}, A. and {Eckart}, A. and {Eisenhauer}, F. and {F{\"o}rster Schreiber}, N.~M. and {Garcia}, P. and {Gao}, F. and {Gendron}, E. and {Genzel}, R. and {Gillessen}, S. and {Habibi}, M. and {Haubois}, X. and {Hei{\ss}el}, G. and {Henning}, T. and {Hippler}, S. and {Horrobin}, M. and {Jim{\'e}nez-Rosales}, A. and {Jochum}, L. and {Jocou}, L. and {Kaufer}, A. and {Kervella}, P. and {Lacour}, S. and {Lapeyr{\`e}re}, V. and {Le Bouquin}, J.-B. and {L{\'e}na}, P. and {Lutz}, D. and {Nowak}, M. and {Ott}, T. and {Paumard}, T. and {Perraut}, K. and {Perrin}, G. and {Pfuhl}, O. and {Rabien}, S. and {Rodr{\'\i}guez-Coira}, G. and {Shangguan}, J. and {Shimizu}, T. and {Scheithauer}, S. and {Stadler}, J. and {Straub}, O. and {Straubmeier}, C. and {Sturm}, E. and {Tacconi}, L.~J. and {Vincent}, F. and {von Fellenberg}, S. and {Waisberg}, I. and {Widmann}, F. and {Wieprecht}, E. and {Wiezorrek}, E. and {Woillez}, J. and {Yazici}, S. and {Young}, A. and {Zins}, G.},
  title   = {Improved {GRAVITY} astrometric accuracy from modeling optical aberrations},
  journal = {Astron. Astrophys.},
  volume  = {647},
  pages   = {A59},
  year    = {2021}
}

@article{Bennett2019,
  author  = {Bennett, M. and Bovy, J.},
  title   = {Vertical waves in the solar neighbourhood in {Gaia} {DR2}},
  journal = {Mon. Not. R. Astron. Soc.},
  volume  = {482},
  pages   = {1417--1425},
  year    = {2019}
}

@article{Reid2020,
  author  = {Reid, M. J. and Brunthaler, A.},
  title   = {The proper motion of {S}agittarius {A}*. {III}. {T}he case for a supermassive black hole},
  journal = {Astrophys. J.},
  volume  = {892},
  pages   = {39},
  year    = {2020}
}

@article{ForemanMackey2019,
  author  = {{Foreman-Mackey}, Daniel and {Farr}, Will and {Sinha}, Manodeep and {Archibald}, Anne and {Hogg}, David and {Sanders}, Jeremy and {Zuntz}, Joe and {Williams}, Peter and {Nelson}, Andrew and {de Val-Borro}, Miguel and {Erhardt}, Tobias and {Pashchenko}, Ilya and {Pla}, Oriol},
  title   = {emcee v3: A {P}ython ensemble sampling toolkit for affine-invariant {MCMC}},
  journal = {J. Open Source Softw.},
  volume  = {4},
  pages   = {1864},
  year    = {2019}
}

@article{Naidu2021,
  author  = {{Naidu}, Rohan P. and {Conroy}, Charlie and {Bonaca}, Ana and {Zaritsky}, Dennis and {Weinberger}, Rainer and {Ting}, Yuan-Sen and {Caldwell}, Nelson and {Tacchella}, Sandro and {Han}, Jiwon Jesse and {Speagle}, Joshua S. and {Cargile}, Phillip A.},
  title   = {Reconstructing the last major merger of the {M}ilky {W}ay with the {H3} {S}urvey},
  journal = {Astrophys. J.},
  volume  = {923},
  pages   = {92},
  year    = {2021}
}

@article{MunozCuartas2011,
  author  = {Mu{\~n}oz-Cuartas, J. C. and Macci{\`o}, A. V. and Gottl{\"o}ber, S. and Dutton, A. A.},
  title   = {The redshift evolution of {$\Lambda$} cold dark matter halo parameters: concentration, spin and shape},
  journal = {Mon. Not. R. Astron. Soc.},
  volume  = {411},
  pages   = {584--594},
  year    = {2011}
}

@article{Dehnen2000,
  author  = {Dehnen, W.},
  title   = {A very fast and momentum-conserving tree code},
  journal = {Astrophys. J. Lett.},
  volume  = {536},
  pages   = {L39--L42},
  year    = {2000}
}

@article{Dehnen2002,
  author  = {Dehnen, W.},
  title   = {A hierarchical {$O(N)$} force calculation algorithm},
  journal = {J. Comput. Phys.},
  volume  = {179},
  pages   = {27--42},
  year    = {2002}
}

@article{Villalobos2008,
  author  = {Villalobos, {\'A}. and Helmi, A.},
  title   = {Simulations of minor mergers---{I}. {G}eneral properties of thick discs},
  journal = {Mon. Not. R. Astron. Soc.},
  volume  = {391},
  pages   = {1806--1827},
  year    = {2008}
}

@misc{Batrakov2026,
  author        = {{Batrakov}, Kirill and {Deason}, Alis J. and {Fragkoudi}, Francesca and {Tomlinson}, Thomas and {Fattahi}, Azadeh and {Belokurov}, Vasily},
  title         = {Slow stellar halo rotation as a signature of disc flips and {GES}-like mergers},
  year          = {2026},
  eprint        = {2609.01208},
  archivePrefix = {arXiv}
}

\end{document}